\documentclass[twocolumn,aps,nofootinbib]{revtex4-1}
\usepackage{amssymb}
\usepackage{graphicx}
\usepackage{amsmath}
\usepackage{color}
\usepackage{subfigure}
\usepackage{mathrsfs}

\usepackage{bm}

\usepackage{times}
\usepackage{amsmath}
\usepackage{amsfonts}
\usepackage{amssymb}
\usepackage{bm}
\usepackage{graphicx}
\usepackage{epstopdf}
\usepackage{enumerate}
\usepackage{color}

\graphicspath{{./figs/}{./}}

\begin{document}
\title{Generalized Flory Theory for Rotational Symmetry Breaking of Complex Macromolecules}
\author{Josh Kelly$^{1}$, Alexander Y. Grosberg$^{2}$, and Robijn Bruinsma$^{1,3}$}
\affiliation{$^{1}$Department of Physics and Astronomy, University of California, Los Angeles, CA 90095, USA}
\affiliation{$^{2}$Department of Physics and Center for Soft Matter Research, New York University, 726 Broadway, New York, NY 10003}
\affiliation{$^{3}$Department of Chemistry and Biochemistry, University of California, Los Angeles, CA 90095, USA}

\begin{abstract}
We report on spontaneous rotational symmetry breaking in a minimal model of complex macromolecules with branches and cycles. The transition takes place as the strength of the self-repulsion is increased. At the transition point, the density distribution transforms from isotropic to anisotropic. We analyze this transition using a variational mean-field theory that combines the Gibbs-Bogolyubov-Feynman inequality with the concept of the \textit{Laplacian matrix}. The density distribution of the broken symmetry state is shown to be determined by the eigenvalues and eigenvectors of this Laplacian matrix. Physically, this reflects the increasing role of the underlying topological structure in determining the density of the macromolecule when repulsive interactions generate internal \textit{tension} Eventually, the variational free energy landscape develops a complex structure with multiple competing minima.
\end{abstract}
\maketitle

It is well known that when attractive interactions between the units (``monomers") of a flexible macromolecule become sufficiently strong, the molecule can undergo a folding transition from a disordered isotropic state to an ordered structure with a specific shape~\cite{pande}. Less familiar is the fact that when repulsive interactions dominate, macromolecules with more complex topologies also can adopt distinct shapes. Examples are dendrimers~\cite{Boris} and certain biopolymers \cite{ding, zipper, gapsys, AjaykumarGopal2012}. Their shape is determined by competing effects. On the one hand, the combination of thermal fluctuations and short-range repulsive interactions between the monomers favors isotropic swelling, since that maximizes the entropy of the molecule. If, however, the topology of the macromolecule constrains the swelling then this generates internal \textit{tension} along the bonds and this reduces the dominance of thermal fluctuations and entropy. Swollen polymer gels~\cite{RedBook} and polymer brushes in good solvent \cite{milner} are familiar examples of polymeric systems where swelling induces a tension that both suppresses fluctuations and confers distinct shape. The suppression of thermal fluctuations means that tense macromolecules of this type can be described by mean-field theory~\cite{milner}, as opposed to linear polymers that have no internal tension~\cite{RedBook}.

Suppose one gradually increases the strength of the repulsions in a soluble macromolecule with a complex topology, is there a well-defined threshold where the molecule develops a distinct shape? If there is such a threshold then what is the nature of the rotational symmetry-breaking transition and how is the resulting shape related to the underlying topology? Finally, if the number of topological constraints is increased, does a complex macromolecule eventually become over-constrained and ``frustrated'' with a free energy landscape that has multiple competing minima \footnote{We focus here on macromolecules with specific, prescribed structures, which is the case of interest for biomolecules. For a discussion of quenched or annealed averages over a class of structures, see refs. \cite{Grosberg1, kantor, grosberg1995}}?

In this paper we propose a theory for the development of shape of topologically complex macromolecules for a minimal model that was introduced by Edwards to describe linear polymers and polymer gels \cite{ball,deam} in good solvent (i.e., solutions where repulsive interactions dominate). We construct a generalization of Flory mean-field theory and apply this to the Edwards Hamiltonian. We find that the density distribution of complex branched polymers indeed undergoes a transition where it loses rotational symmetry. The structure of the broken symmetry state is determined by the eigenvalues and eigenvectors of the \textit{Laplacian matrix} of the molecule, a concept borrowed from graph theory. As the strength of the repulsive interactions further increases, a complex energy landscape emerges with multiple competing minima. We find that at least the coarse-grained features of the density distribution of complex macromolecules and the tension profile can be predicted on the basis of the eigenvalues and eigenvectors of the Laplacian matrix.

The Edwards Hamiltonian for a macromolecule is defined by
\begin{equation}
\beta H = \frac{d}{2a^2} {\sum_{i<j}}^{\prime}({\bf{r}}_i - {\bf{r}}_j)^2+ \sum_{i<j}u(|{\bf{r}}_i - {\bf{r}}_j|)
\label{ch3:H2}
\end{equation}
The summations are here over $N$ point-like monomers located at sites ${\bf{r}}_i$ with $i=1,2,..N$ that are linked into a connected network by harmonic springs. The prime in $H_0$, the first term on the right-hand side, indicates that this double sum is to be restricted to monomers pairs that are linked by springs. The second term in Eq.\ref{ch3:H2} represents short-range repulsive monomer-monomer interactions with strength $v=\int u(r) d^dr$ and range $\sigma$ in units of $a$. The summation now is over all monomer pairs. The Edwards Hamiltonian is realized by a network of cross-linkers connected by ideal polymer chains that have an RMS radius of gyration $a$.

Assume that the radius of gyration $R_0$ of the molecule for $v=0$ has been determined (we will shortly see how). The radius of gyration for $v\neq 0$ can then be obtained by minimizing the Flory variational free energy \cite{RedBook}
\begin{equation}
\begin{split}
&\beta F_F(R) =\left(\frac{R^2}{R_0^2}\right)+v \frac{N^2}{R^d}
\label{Flory}
\end{split}
\end{equation}
(dropping numerical coefficients). The first term on the right hand side represents entropic elasticity resisting the swelling while the second term represents osmotic swelling pressure due to monomer-monomer repulsion, as expressed in second viral form. Minimizing $F_F(R)$ with respect to $R^2$ leads to the familiar result that $(R/R_0)^2$ increases as $(v N^2/R_0^3)^{2/5}$ in $d=3$ when the strength of the repulsion increases. Flory theory implicitly assumes a uniform and isotropic density.

In order to allow for an anisotropic density, we first recast the Flory variational energy as a special case of the Gibbs-Bogolyubov-Feynman (GBF) variational principle~\cite[Section 1]{SupMat}, which states that
\begin{equation}
F \leq F_T + \left<(H - H_T)\right>_T
\end{equation}
Here, $\left< \ldots \right>_T$ indicates that a Boltzmann average is to be taken with respect to the trial Hamiltonian $H_T$. $F_T$ is the free energy associated with $H_T$. The variational free energy $F_V=F_T + \left< (H - H_T) \right>_T$ provides an upper bound for the free energy.

For $H_T$, we will use generalized versions of $H_0$ expressed in terms of the eigenvectors and eigenvalues of the $N\times N$ real, square, symmetric Laplacian matrix $L_{i,j}$. The Laplacian matrix is the Laplace operator in matrix form defined on a graph of the nodes and bonds of the molecule. It has been extensively studied in the context of graph theory \cite{LaplacianNotes}.  Diagonal entries $L_{n,n}$ are equal to the number of monomers linked to monomer $n$ (``vertex degree") while off-diagonal entries $L_{n,m}$ are equal to $-1$ if monomer $n$ and $m$ are linked and $0$ if they are not. The rows and columns of $L_{i,j}$ add to zero so the $N$ component vector with entries equal to one is an eigenvector with eigenvalue $0$. The other eigenvalues $\lambda^{(j)}$, with $j=1,2,...N-1$, are strictly positive for a connected graph. The lowest non-zero eigenvalue $\lambda^{(1)}$, henceforth denoted by $\lambda$, is known as the ``spectral gap'' \footnote{Analytical expressions for the eigenvalues are available for linear chains, cubic lattices, dendrimers, and a variety of fractal structures \cite{doi:10.1063/1.4794921,dolgushev2016extended, julaiti}. Efficient algorithms are available for the numerical computation of the eigenvalues and eigenvectors.}. Note that the eigenvalues and eigenvectors of the Laplacian matrix reflect only the topology of the graph of the molecule and do not relate to the geometrical space in which the molecule is embedded.

In terms of the Laplacian matrix $H_0$ can be written by unrestricted summation over all particles in the form $\beta H_0 =\left(\frac{d}{2a^2}\right){\sum\limits_{i,j}}L_{i,j}{\mathbf{r}_i}\cdot{{\mathbf{r}}_j}$. This can be usefully expressed in the form $\beta H_0 =\frac{d}{ 2}\sum\limits_{j=1}^{N-1}\lambda^{(j)}|{\bf{A}}^{(j)}|^2$ where the $\lambda^{(j)}$ are the (rank-ordered) eigenvalues of the Laplacian matrix and where the ${\bf{A}}^{(j)}=\sum\limits_{i=1}^{N-1}{\bf{r}}_{i}\xi_i^{(j)}/a$ are the normal mode amplitudes. The latter are vectors in the $d$-dimensional embedding space but expressed in terms of the orthonormal $N$-component eigenvectors $\xi^{(j)}_j$ of the Laplacian matrix \cite{Nitta1}. The mode amplitudes can be viewed as analogs of the Fourier amplitudes describing the displacements of the nodes of a graph embedded in a $d$-dimensional space \footnote{Examples of eigenvalues and eigenvectors of the Laplacian matrix and of the normal modes are given in~\cite[Section II]{SupMat}}.

The mean square radius of gyration $R_0^{2}$ of the ideal molecule can be expressed in terms of the eigenvalues as $\frac{a^2}{N}\sum_{j=1}^{N-1}{\frac{1}{\lambda^{(j)}}}$ \cite{Nitta1}, which can be viewed as a generalization of the Kramers Theorem \cite{Rubinstein}. If the spectral gap $\lambda$ is small compared to the higher eigenvalues then this reduces to $\lambda\simeq a^2/(NR_0^2) $. \footnote{For the case of the ring polymer in \cite[Section II]{SupMat}, the computation of the mean square radius of gyration reproduces the standard result that $R_0^{2}\propto N$. The ${\bf{A}}^{(1)}$ spectral gap mode corresponds to an expansion of the $N$ particles from a point to a ring with radius $\left| {\bf{A}}^{(1)} \right|$.}. An important unphysical feature of $H_0$ is that it has $d$ zero modes (associated with translation symmetry) whereas the correct number of zero modes of a physical molecule -- including translation and rotation symmetry -- is $d(d+1)/2$ ($3$ in $d = 2$ and $6$ in $d = 3$).

As a first example of the use of the variational method to include monomer-monomer interaction, assume that $H_T$ equals $H_0$ except that the lowest non-zero eigenvalue, the spectral gap $\lambda$, is replaced by a variational parameter $\gamma$. So:
\begin{equation}
\beta H_T =\frac{d}{ 2}\gamma|{\bf{A}}^{(1)}|^2+\frac{d}{ 2}\sum_{i=2}^{N-1}\lambda^{(i)}|{\bf{A}}^{(i)}|^2
\label{H0}
\end{equation}
This leads to a variational free energy:
\begin{equation}
\beta F_V(\gamma) \simeq \frac{d}{2}\left(\ln\gamma+\frac{\lambda}{\gamma}\right) + C(N)\frac{v}{a^d} \left( \frac{\gamma d}{2\pi} \right)^{d/2}
\label{Flory}
\end{equation}
where $C(N)=\sum_{m<n=1}^N \left(({\xi}_m-{\xi}_n)^2+\gamma\sigma^2 \right)^{-d/2}$ with $\xi_m$ the eigenvector associated with the spectral gap $\lambda$ and $\sigma$ the range of the excluded volume interaction in units of $a$ (the derivation is given in \cite[Section III]{SupMat}). Using the normalization $\sum_{m=1}^N{\xi}_m^2=1$ and assuming a (large) random structure gives $C(N)\propto N^{2+d/2}$. The resulting variational expression reduces to Flory mean-field theory if one replaces $\gamma$ by $a^2/(NR^2)$.

Next, allow for the possibility of an anisotropic density by including in $H_T$ non-zero expectation values ${\bf{A}}_{0}^{(i)}$ for the $M$ mode amplitudes with the lowest $M$ eigenvalues:
\begin{equation}
\beta H_T =  \frac{d}{ 2} \left(\sum_{i=1}^M\gamma^{(i)}\left( {\bf{A}}^{(i)}-{\bf{A}}_{0}^{(i)} \right)^2+\sum_{i=M+1}^{N}\lambda^{(i)}|{\bf{A}}^{(i)}|^2\right)
\label{ch3:H4}
\end{equation}
The special case $\gamma^{(i)}=\lambda^{(i)}$ and $M=N$ is interesting.  The set of order-parameters ${\bf{A}}_{0}^i$ then defines a set of $N$ particle vectors ${{\bf{r}}_0}_j/a=\sum_{i=1}^{N-1} {\bf{A}}_{0}^i\xi_j^{(i)}$ (up to an overall translation). Expressing the trial Hamiltonian in real space leads to:
\begin{equation}
\beta H_T = \frac{d}{2a^2}{\sum_{i<j}}^{\prime}({\bf{r}}_i - {\bf{r}}_j-\Delta {{\bf{r}}_0}_{ij})^2
\label{GNM}
\end{equation}
where $\Delta {{{\bf{r}}}_0}_{i,j}=({{\bf{r}}_0}_i-{{\bf{r}}_0}_j)$. This is similar to the Hamiltonian of the ideal molecule except that Gaussian bonds ${\bf{r}}_i - {\bf{r}}_j$ linking monomers $i$ and $j$ have been placed under internal tension so the expectation value of the bond separation $\Delta {{\bf{r}}^0}_{i,j}$ has a certain direction in space. Formally, Eq.\ref{GNM} is identical to the Gaussian Network Model that is frequently used to obtain the normal modes of folded proteins \cite{gaussian}. It also has the appropriate number of zero modes.

Next, include both types of variation. The simplest case is again $M=1$:
\begin{equation}
\begin{split}
\beta F_V(\gamma,{\bf{A}}_0) & \simeq \frac{d}{2}\ln\gamma+\frac{d}{2}\lambda\left(|{\bf{A}}_0|^2+\frac{1}{\gamma}\right) \\& + C(N)\frac{v}{a^d} \left( \frac{\gamma d}{2\pi} \right)^{d/2} e^{ -\gamma d{\bf{A}}_0^2/2}
\label{M=1}
\end{split}
\end{equation}
The function $F_V(\gamma,|{\bf{A}}_0|)$ always has a stable minimum at $|{\bf{A}}_0|=0$, which corresponds to Flory theory, but he surface $F_V(\gamma,{\bf{A}}_0)$ has a second minimum for a large $\gamma$ and a non-zero value of  $|{\bf{A}}_0|$. The density distribution described by this minimum has a stretched, linear shape with particle locations determined by the eigenvector $\xi_m$ of the Laplacian matrix associated with the spectral gap. As a function of increasing $v/a^d$, the absolute minimum usually shifts discontinuously from the Flory minimum to the new minimum. An exception is the case of a linear chain when the Flory minimum is the absolute minimum for any $v$.

\begin{figure}
	\centering
	\includegraphics[width=3.4in]{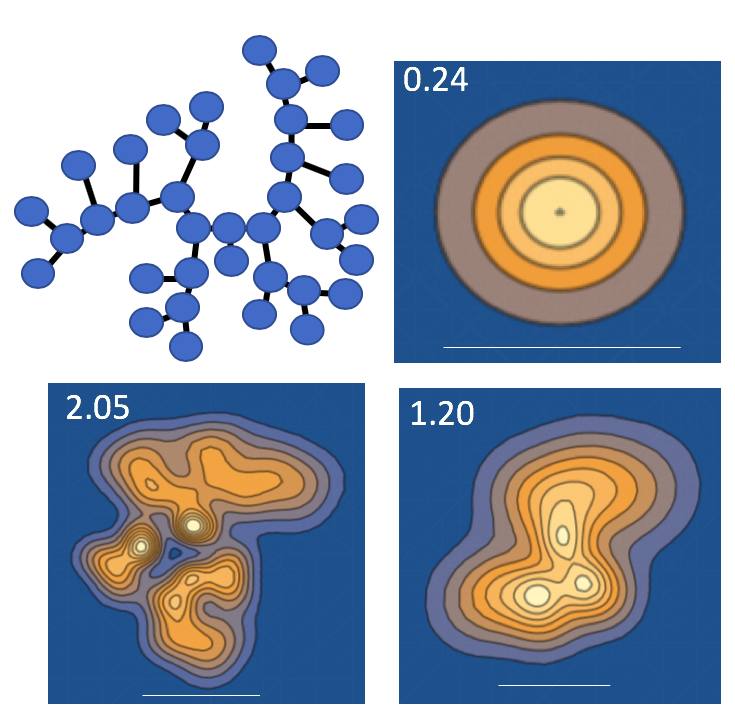}
	\caption{Two-dimensional density profiles obtained using the GBF variational method for a 36 node branched polymer with a maximum of $M=18$ non-zero mode expectation values. Top left: Graph of the molecule. Top right: $v/a^2 = 0.24$ in units of $k_BT$. The density profile is isotropic.
Bottom right: $v/a^2 =1.20$. A few low-lying modes have condensed. Bottom left: $v/a^2 =2.05$. Most modes have condensed. White space bar: $5a$}.
	\label{MCS}
\end{figure}

While the $M=1$ theory is analytically tractable, it can be shown that it only allows for linearly stretched shapes and that it is necessary to include multiple modes to completely lift the overlap between particles~\cite[Section I]{SupMat}. The variational free energy $F_V(\{\gamma^{(i)},{\bf{A}}_0^{(i)}\})$ for $M$ coupled vectorial order parameters is a natural extension of Eq. \ref{M=1} (derived in~\cite[Section III]{SupMat}) but minimization of $F_V(\{\gamma^i,{\bf{A}}_0^{(i)}\})$ requires numerical methods. Numerical minimization of $F_V(\{\gamma^i,{\bf{A}}_0^{(i)}\})$ for a second-generation dendrimer in $d=2$ showed that for increasing $v/a^d$, there is now a whole series of transitions where modes with increasing eigenvalues ``freeze out''. Importantly, the interacting system has the correct number of zero modes in $d=2$. The numerical minimization of $F_V(\{\gamma^i,{\bf{A}}_0^{(i)}\})$ for a 36-node branched graph with a maximum of $M=18$ mode expectation values is shown in Fig.~\ref{MCS}.
For $v/a^2$ less than about $0.75$, the isotropic Flory minimum was the lowest free energy state, as illustrated by the case $v/a^2 = 0.24$. However, for $v/a^2=1.20$ the density profile is quite anisotropic with three diffuse maxima. The power spectrum of mode amplitudes in this state is dominated by the lowest few eigenvalues. For $v/a^3=2.05$, all of the $M=18$ modes have gained non-zero expectation values and the power spectrum is more complex with a second peak at larger eigenvalues. Note that the density profile is quite detailed. The system appears to be frozen. However, the numerical minimization of the variational free energy was, for larger values of $v/a^2$, significantly complicated by the fact that the variational free energy clearly had numerous minima with comparable energies. The last density profile should be viewed only as representative.

\begin{figure}
	\centering
	\includegraphics[width=3.0in]{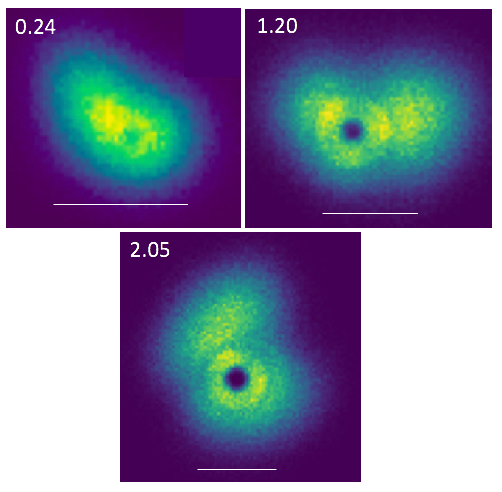}
	\caption{Density profiles obtained by Monte-Carlo simulation for the same molecule and interaction strengths as Fig. \ref{MCS}. White space bar: $5a$. }.
	\label{ch3:d20}
\end{figure}

For comparison, we also performed a $d=2$ Monte-Carlo (MC) simulations on the same system (Fig.\ref{ch3:d20}).
One of the nodes was pinned to suppress rigid-body Brownian motion of the center of mass. The Kabsch algorithm \cite{kabsch} was used to compensate for rigid-body rotational Brownian motion. The top left image in Fig.\ref{ch3:d20}, with $v/a^2 = 0.24$, has a radius of gyration comparable to the theoretical prediction and a weak but noticeable rotational asymmetry. For  $v/a^2 = 1.20$, the predicted and computed densities have comparable sizes and both have three maxima. The onset of rotational asymmetry thus appears to be less sharp than predicted by the theory while for $v/a^2 = 2.05$ the theoretical density profile is significantly more detailed than the computed profile. The MC simulations were, in this last case, complicated by long relaxation times.

As an alternative route for a quantitative test of the theory, we compared the moduli $|\Delta {{{\bf{r}}}_0}_{i,j}|$ of the bond extensions predicted by the GBF variational principle with those obtained from the MC simulation. We found that the variational method correctly obtains the bond extensions of the outer monomers while it somewhat overestimates the bond extensions of the inner monomers (\cite[Section IV]{SupMat}). Note that the development of significant internal tension provides an \textit{a-posteriori} justification of the use of a self-consistent mean-field theory. The agreement in terms of the bond tensions but not in terms of detailed density profiles indicates that the competing free energy minima may have similar patterns of bond tension. An important extension of the theory would be to allow for the fact that this is a finite system with multiple minima computing to the free energy. This would lead to ``smearing" of the rotational symmetry breaking transition.

A natural area where this theory can be applied is that of biopolymers with non-trivial topology that are dominated by repulsive interactions. An increasing number of functional but disordered proteins has been identified. The interactions between the different parts of these proteins are predominantly repulsive (``good solvent'') yet they have distinct, reproducible shapes \cite{ding, zipper}, as confirmed by Molecular Dynamics simulations~\cite{gapsys}. Though proteins have a linear polymer primary structure, they still can effectively adopt a complex topology due to attractive interactions between specific residues, for instance between cysteine residues that can form disulfide bridges. Another possible area of application involves the shape of large, single-stranded RNA molecules. A graph of the secondary structure of an RNA molecule has a branched topology without circuits, The tertiary structure of an RNA molecule is generated by pairing between non-adjacent nucleic acids that were not paired as part of the secondary structure and these tertiary contacts could be included as bonds in the graph of the molecule, which would produce cycles. Cryo-EM studies of large, swollen single-stranded RNA molecules in good solvents reveal that they are disordered but their density profile has a distinct anisotropy~\cite{gopal2011}. Current methods of DNA origami allow for the construction of molecular structures with prescribed topologies that are reasonably represented by the Edwards Hamiltonian, which allows for direct experimental tests of the proposed theory. A rotational symmetry breaking transition could be engineered by changing the solvent quality. We close by noting that there is a related problem where the method discussed in this paper could be applied namely the computation of the most likely structure of a biopolymer for which it already has been experimentally determined that certain elements of the primary structure are adjacent to each other, for example in the form of a contact map obtained by NMR \cite{gutin1994}. Though the distance constraints are here \textit{knowledge-based}, instead of physical bonds or links, the Laplacian matrix method still could be used to encode the NMR contact map after which likely density profiles could be computed using the method we outlined here.

\acknowledgements
RB thanks Alex Levine for useful discussion and acknowledges support from NSF-DMR under Grant 1006128 and from the Simons Foundation.  Both A.Y.G. and R.B. wish to acknowledge the Aspen Center for Physics supported by the National Science Foundation (USA) under Grant No. PHY-1066293 where this work was started.

\bibliography{thesis2}

\end{document}